\documentclass[12pt]{article}%
\usepackage{amsmath,latexsym}
\usepackage{graphicx}
\usepackage{amsmath}
\usepackage{amsfonts}
\usepackage{amssymb}%
\begin{document}

\title{{Traversable wormholes sustained by an extra 
   spatial dimension}}
   \author{
Peter K. F. Kuhfittig*\\  \footnote{kuhfitti@msoe.edu}
 \small Department of Mathematics, Milwaukee School of
Engineering,\\
\small Milwaukee, Wisconsin 53202-3109, USA}

\date{}
 \maketitle

\begin{abstract} This paper explores the effect of an extra
spatial dimension on a Morris-Thorne wormhole.  After
proposing a suitable model, it is shown that under certain
conditions, the throat of the wormhole can be threaded with
ordinary matter, while the unavoidable violation of the null
energy condition can be attributed to the existence of the
extra dimension.
\end{abstract}

\maketitle

\vspace{12pt}

\section{Introduction}
Wormholes are handles or tunnels connecting widely separated
regions of our Universe or entirely different universes.
That such structures might be suitable for interstellar
travel was first proposed by Morris and Thorne \cite{MT88}.
Holding a wormhole open requires the use of ``exotic matter."
Such matter violates the null energy condition, discussed
further in Sec. \ref{S:structure}.

While wormholes are valid predictions of Einstein's theory,
quantum field theory places severe restrictions on their
existence \cite{FR1, FR2, Ford1, Ford2}.  For example, Ford
and Roman \cite{Ford1, Ford2} have shown that the wormholes
in Ref. \cite{MT88} could not exist on a macroscopic scale.
The wormholes in Refs. \cite{MTY} and \cite{pK09} could
exist but are subject to extreme fine-tuning in order to
minimize the amount of exotic matter.  On the positive side,
it has been shown that phantom dark energy violates the null
energy condition and could in principle be used for wormhole
construction \cite{sS05, fL05}.

Given the problematical nature of exotic matter in an ordinary
Morris-Thorne wormhole, other approaches have been undertaken.
For example, it was proposed by Lobo and Oliveira \cite{LO09}
that in the context of $f(R)$ modified gravity, a wormhole
could be constructed from ordinary matter, while the violation
of the null energy condition can be attributed to the effect
of the modified gravity theory.  Noncommutative geometry, an
offshoot of string theory, also offers a way for allowing
ordinary matter since, analogously, the violation of the
null energy condition is due to the noncommutative-geometry
background \cite{fR12, pK15}.  Finally, Ref. \cite{pK18}
discusses the combined effects of $f(R)$ gravity and
noncommutative geometry.

In this paper we offer yet another approach by
hypothesizing the existence of an extra spatial dimension.
Any proposal dealing with an extra spatial dimension requires
a look backward.  An extension of Einstein's general theory
of relativity from four to five dimensions by T. Kaluza was
intended to be a major step toward a unified field theory, the
combination of Einstein's and Maxwell's theories.  The idea was
later extended to include the nuclear forces, leading to several
more (compactified) dimensions to become string/M theory.

A strong advocate of retaining an extra spatial dimension has
been Paul Wesson \cite{WP92}.  One reason is that the field
equations for a five-dimensional totally flat space yield
Einstein's equations in four dimensions containing matter,
also called the \emph{induced-matter theory}.  It can be
argued that our understanding of four-dimensional gravity,
including the equivalence principle, is greatly enhanced by
assuming a fifth dimension \cite{pW15, pW13}.  Unfortunately,
merely introducing a fifth dimension tells us nothing about
its form or its physical nature.  Our first task must
therefore be the construction of a model that is consistent
with our knowledge of general relativity.  What naturally
comes to mind is a static and spherically symmetric form in
Schwarzschild coordinates: given that in four dimensions
we have the line element
\begin{equation}\label{E:L1}
  ds^2=-e^{2\Phi(r)}dt^2 +e^{2\lambda(r)}dr^2 +r^2(d\theta^2
  +\text{sin}^2\theta\,d\phi^2),
\end{equation}
a fifth term may well have the form $e^{2\mu(r,l)}dl^2$,
where $l$ is the extra coordinate.  It should be noted that we
are retaining the dependence on the $r$-coordinate to make the
model as general as possible.  So the line element becomes
\begin{equation}\label{E:L2}
  ds^2=-e^{2\Phi(r)}dt^2 +e^{2\lambda(r)}dr^2+r^2
  (d\theta^2+\text{sin}^2\theta\,d\phi^2)+e^{2\mu(r,l)}dl^2,
\end{equation}
where we assume that $c=G=1$.

Our main goal is to show that in the four-dimensional setting,
the throat of the wormhole can be threaded with ordinary
matter, while the unavoidable violation of the null energy
condition can be attributed to the existence of the extra
dimension.

\emph{Remark:} In line element (\ref{E:L2}), the functions
$\Phi$ and $\lambda$ could conceivably be functions of both $r$
and $l$, thereby matching the form of $\mu(r,l)$.  We will
return to this case in Sec. \ref{S:remaining}.

\section{Wormhole structure}\label{S:structure}
Wormholes may be described by the static and spherically
symmetric line element
\begin{equation}\label{E:L3}
ds^{2}=-e^{2\Phi(r)}dt^{2}+\frac{dr^2}{1-b(r)/r}
+r^{2}(d\theta^{2}+\text{sin}^{2}\theta\,
d\phi^{2}),
\end{equation}
again using units in which $c=G=1$.  Here $\Phi =\Phi(r)$ is
called the \emph{redshift function}, which must be everywhere
finite to prevent an event horizon.  The function $b=b(r)$ is
called the \emph{shape function}.  (It determines the spatial
shape when viewed in the usual embedding diagram \cite{MT88}.)
The spherical surface $r=r_0$ is the \emph{throat} of the
wormhole.  The shape function must satisfy the following
conditions: $b(r_0)=r_0$, $b(r)<r$ for $r>r_0$, and $b'(r_0)
\le 1$, called the \emph{flare-out condition} in Ref. \cite
{MT88}.  For a Morris-Thorne wormhole, these conditions can
only be satisfied by violating the null energy condition
(NEC), which states that for the energy-momentum tensor
$T_{\alpha\beta}$,
\begin{equation}\label{E:null}
   T_{\alpha\beta}k^{\alpha}k^{\beta}\ge 0\,\,
   \text{for all null vectors}\,\, k^{\alpha}.
\end{equation}
For example, given an orthonormal frame, $T_{00}=\rho$
is the energy density and $T_{11}=p_r$ is the radial
pressure; then the outgoing radial null vector
$(1,1,0,0)$ yields $\rho +p_r<0$ whenever the NEC
is violated.  Matter that violates the NEC is called
``exotic" in Ref. \cite{MT88}.  (These ideas will be
discussed further in Sec. \ref{S:remaining}.)

Returning to Eq. (\ref{E:L2}), for the line element in
this paper, we assume that $e^{2\lambda(r)}=1-b(r)/r$.  Thus
\begin{equation}\label{E:L4}
  ds^2=-e^{2\Phi(r)}dt^2+\frac{dr^2}{1-b(r)/r}+r^2
  (d\theta^2+\text{sin}^2\theta\,d\phi^2)+e^{2\mu(r,l)}dl^2.
\end{equation}

As noted in the Introduction, the main goal in this paper is
to determine the effect that the extra spatial dimension may
have on a (four-dimensional) Morris-Thorne wormhole.  More
precisely, we are going to show that under certain conditions,
the NEC is satisfied within the four-dimensional framework.
However, the same conditions also lead to a null vector in
the five-dimensional spacetime for which the NEC is violated.
So while the wormhole can be constructed from ordinary matter,
it is sustained by the existence of the higher dimension,
provided, of course, that certain conditions are met.

\section{The solution}
To study the effect of the extra spatial dimension, knowledge
of classical Morris-Thorne wormholes is of little help.
So we need to fall back on certain basic principles.  To
that end, we choose an orthonormal basis
$\{e_{\hat{\alpha}}\}$ which is dual to the following 1-form basis:
\begin{multline}\label{E:oneform1}
    \theta^0=e^{\Phi(r)}\, dt,\quad \theta^1=[1-b(r)/r]^{-1/2}\,dr,\\
     \quad\theta^2=r\,d\theta, \quad
      \theta^3=r\,
\,\text{sin}\,\theta\,d\phi,\quad \theta^4=e^{\mu(r,l)}dl.
\end{multline}
These forms yield
\begin{multline}\label{E:oneform3}
     dt=e^{-\Phi(r)}\,\theta^0,\quad dr=[1-b(r)/r]^{1/2}\,\theta^1,\\
     \quad d\theta=\frac{1}{r}\theta^2, \quad
    d\phi=\frac{1}{r\,\text{sin}\,\theta}\theta^3, \quad
    dl=e^{-\mu(r,l)}\,\theta^4.
\end{multline}
To obtain the curvature 2-forms and the components of the
Riemann curvature tensor, we use the method of differential
forms, following Ref. \cite{HT90}.  So the next step is to
calculate the following exterior derivatives in terms of
$\theta^i$, where $b=b(r)$:
\begin{equation}
   d\theta^0=\frac{d\Phi(r)}{dr}\left(1-\frac{b}{r}\right)
   ^{1/2}\,\theta^1\wedge\theta^0, \quad d\theta^1=0,
\end{equation}
\begin{equation}
    d\theta^2=\frac{1}{r}\left(1-\frac{b}{r}\right)^{1/2}
    \theta^1\wedge\theta^2,
\end{equation}
\begin{equation}
     d\theta^3=\frac{1}{r}\left(1-\frac{b}{r}\right)^{1/2}
     \theta^1\wedge\theta^3+\frac{1}{r}\text{cot}\,\theta
     \,\,\theta^2\wedge\theta^3,
\end{equation}
\begin{equation}
   d\theta^4=\frac{\partial \mu(r,l)}{\partial r}
   \left(1-\frac{b}{r}\right)^{1/2}
   \,\theta^1\wedge\theta^4.
\end{equation}
The connection 1-forms $\omega^i_{\phantom{i}\,\,k}$ have the
symmetry
    $\omega^0_{\phantom{i}\,\,i}=\omega^i_{\phantom{0}0}
    \;(i=1,2,3,4)$\;\text{and}\;$\omega^i_{\phantom{j}j}=
     -\omega^j_{\phantom{i}\,i}\;(i,j=1,2,3,4, i\ne j)$,
and are related to the basis $\theta^i$ by
\begin{equation}
   d\theta^i=-\omega^i_{\phantom{k}k}\wedge\theta^k.
\end{equation}
The solution of this system is found to be
\begin{equation}
   \omega^0_{\phantom{0}1}=\frac{d\Phi(r)}{dr}
   \left(1-\frac{b}{r}\right)^{1/2}\theta^0,
\end{equation}
\begin{equation}
   \omega^2_{\phantom{0}1}=
   \frac{1}{r}\left(1-\frac{b}{r}\right)^{1/2}\theta^2,
\end{equation}
\begin{equation}
    \omega^3_{\phantom{0}1}=
    \frac{1}{r}\left(1-\frac{b}{r}\right)^{1/2}\theta^3,
\end{equation}
\begin{equation}
   \omega^3_{\phantom{0}2}=
   \frac{1}{r}\,\text{cot}\,\theta\,\,\theta^3,
\end{equation}
\begin{equation}\label{E:omega}
   \omega^4_{\phantom{0}1}=
   \frac{\partial\mu(r,l)}{\partial r}
      \left(1-\frac{b}{r}\right)^{1/2}\theta^4,
\end{equation}
\begin{equation}
   \omega^0_{\phantom{0}2}=\omega^0_{\phantom{0}3}=
   \omega^0_{\phantom{0}4}=\omega^2_{\phantom{0}4}=
   \omega^3_{\phantom{0}4}=0.
\end{equation}

The curvature 2-forms $\Omega^i_{\phantom{j}j}$ are calculated
directly from the Cartan structural equations
\begin{equation}
    \Omega^i_{\phantom{j}j}=d\omega^i_{\phantom{j}j} +\omega^i
     _{\phantom{j}k}\wedge\omega^k_{\phantom{j}j}.
\end{equation}
The results for $\Omega^i_{\phantom{j}j}$ are given in
Appendix A.

The components of the Riemann curvature tensor can be read off
directly from the form
\begin{equation}
   \Omega^i_{\phantom{j}j}=-\frac{1}{2}R_{mnj}^{\phantom{mnj}i}
    \;\theta^m\wedge\theta^n
\end{equation}
and are listed next:
\begin{equation}
   R_{011}^{\phantom{000}0}=-\frac{1}{2}\frac{d\Phi(r)}{dr}
   \frac{rb'-b}{r^2}+\frac{d^2\Phi(r)}{dr^2}\left(1-\frac{b}{r}
   \right)
   +\left[\frac{d\Phi(r)}{dr}\right]^2\left(1-\frac{b}{r}\right),
\end{equation}
\begin{equation}
   R_{022}^{\phantom{000}0}=R_{033}^{\phantom{000}0}=
   \frac{1}{r}\frac{d\Phi(r)}{dr}\left(1-\frac{b}{r}\right),
\end{equation}
\begin{equation}
   R_{044}^{\phantom{000}0}=\frac{d\Phi(r)}{dr}
     \frac{\partial\mu(r,l)}{\partial r}\left(1-\frac{b}{r}\right),
\end{equation}
\begin{equation}
   R_{122}^{\phantom{000}1}=R_{133}^{\phantom{000}1}=
   -\frac{1}{2}\frac{rb'-b}{r^3},
\end{equation}
\begin{equation}
   R_{144}^{\phantom{000}1}=\frac{\partial^2\mu(r,l)}{\partial r^2}
   \left(1-\frac{b}{r}\right)-\frac{1}{2}\frac{\partial\mu(r,l)}
   {\partial r}\frac{rb'-b}{r^2}
   +\left[\frac{\partial\mu(r,l)}{\partial r}\right ]^2
   \left(1-\frac{b}{r}\right),
\end{equation}
\begin{equation}
   R_{233}^{\phantom{000}2}=-\frac{b}{r^3},
\end{equation}
\begin{equation}
   R_{244}^{\phantom{000}2}=R_{344}^{\phantom{000}3}=
   \frac{1}{r}\frac{\partial\mu(r,l)}{\partial r}
      \left(1-\frac{b}{r}\right).
\end{equation}

The last form to be derived in this section is the Ricci
tensor, which is obtained by a trace on the Riemann
curvature tensor:
\begin{equation}
   R_{ab}=R_{acb}^{\phantom{000}c}.
\end{equation}
The components are listed in Appendix B.

\section{The main result}
First we recall from Sec. \ref{S:structure} that the
NEC states that for the energy-momentum tensor
$T_{\alpha\beta}$,
$T_{\alpha\beta}k^{\alpha}k^{\beta}\ge 0\,\,
\text{for all null vectors}\,\, k^{\alpha}$.  Furthermore,
an ordinary Morris-Thorne wormhole can only be maintained
if this condition is violated, thereby requiring exotic
matter.   We wish to show in this section that thanks to
the extra spatial dimension, the wormhole throat can be
threaded by ordinary matter and that the violation of
the NEC can be attributed to the existence of the fifth
dimension.

Let us start with the four-dimensional null vector
$(1,1,0,0)$, leaving the other null vectors for later.
The Einstein field equations in the orthonormal frame
are
\begin{equation}
   G_{\hat{\alpha}\hat{\beta}}=R_{\hat{\alpha}\hat{\beta}}-\frac{1}
{2}Rg_{\hat{\alpha}\hat{\beta}}=8\pi T_{\hat{\alpha}\hat{\beta}},
\end{equation}
where
\begin{equation}
   g_{\hat{\alpha}\hat{\beta}}=
   \left(
   \begin{matrix}
   -1&0&0&0\\
   \phantom{-}0&1&0&0\\
   \phantom{-}0&0&1&0\\
   \phantom{-}0&0&0&1
   \end{matrix}
   \right).
\end{equation}
To avoid this rather cumbersome notation and to be
consistent with the previous section, we will now
omit the hats.  So $T_{00}=\rho$ is the energy
density and $T_{11}=p_r$ is the radial pressure, as
noted after Eq. (\ref{E:null}).  Hence
\begin{equation}\label{E:exotic}
   8\pi (\rho +p_r)=[R_{00}-\frac{1}{2}R(-1)]+[R_{11}-
   \frac{1}{2}R(1)]=R_{00}+R_{11}.
\end{equation}
This important special case says that the radial tension
$\tau=-p_r$ exceeds $\rho c^2$ whenever the NEC is violated
and essentially characterizes exotic matter.  Since we are
primarily interested in the vicinity of the throat, we
assume that $1-b(r_0)/r_0=0$.  Making use of Appendix B,
Eq. (\ref{E:exotic}) then yields
\begin{equation}
  \left.\rho +p_r\right|_{r=r_0}
  =\frac{1}{8\pi}\frac{b'(r_0)-1}{r_0^2}
  \left[1+\frac{r_0}{2}\frac{\partial\mu(r_0,l)}{\partial r}
  \right].
\end{equation}
Recalling that $b'(r_0)<1$, the flare-out condition, we
obtain
\begin{equation}
  \rho+p_r>0 \quad  \text{at} \quad r=r_0
\end{equation}
provided that
\begin{equation}\label{E:condition1}
   \frac{\partial\mu(r_0,l)}{\partial r}<-\frac{2}{r_0}.
\end{equation}
It is interesting to note that if $\mu(r,l)$ is independent of
$r$, so that $\partial\mu(r,l)/\partial r=0$, then
\begin{equation*}
   \left.\rho +p_r\right|_{r=r_0}
  =\frac{1}{8\pi}\frac{b'(r_0)-1}{r_0^2}<0,
\end{equation*}
the usual result for a Morris-Thorne wormhole.

Condition (\ref{E:condition1}) also plays a role in the
violation of the NEC in the higher-dimensional space.
To show this, consider the null vector $(1,0,0,0,1)$.
Assuming that the Einstein field equations hold in the
five-dimensional space, we now have
\begin{equation*}
  G_{00}+G_{44}=(R_{00}-\frac{1}{2}Rg_{00})+
  (R_{44}-\frac{1}{2}Rg_{44})=R_{00}+R_{44}
\end{equation*}
and, again assuming that $1-b(r_0)/r_0=0$,
\begin{equation}\label{E:R4}
    \left.G_{00}+G_{44}\right|_{r=r_0}=\\
    \frac{1}{2}\frac{rb'-b}{r^2}\left[
    -\frac{d\Phi(r)}{dr}+\frac{\partial\mu(r,l)}
    {\partial r}\right]_{r=r_0}.
\end{equation}
Since $\partial\mu(r_0,l)/\partial r<-2/r_0$, the second
factor on the right side of Eq. (\ref{E:R4}) is positive
if
\begin{equation}\label{E:condition2}
   \frac{d\Phi(r_0)}{dr}=-A<
   \frac{\partial\mu(r_0,l)}{\partial r} <-\frac{2}{r_0},
\end{equation}
similar to Inequality (\ref{E:condition1}).

\section{The remaining conditions}\label{S:remaining}
One physical consequence of the NEC is that it forces the
local energy density to be positive.  To check this
requirement, we need the Ricci scalar
\begin{equation}\label{E:Ricci}
R=R^i_{\phantom{0}i}=-R_{00}+R_{11}+R_{22}+R_{33}+R_{44}.
\end{equation}
 As already noted, $8\pi\rho=R_{00}+\frac{1}{2}R$.  It
 is readily checked that
 \begin{equation}
    G_{00}=R_{00}+\frac{1}{2}R=-R_{122}^{\phantom{000}1}
    -R_{133}^{\phantom{000}1}-R_{144}^{\phantom{000}1}
    -R_{233}^{\phantom{000}2}-R_{244}^{\phantom{000}2}
    -R_{344}^{\phantom{000}3}.
 \end{equation}
So at $r=r_0$, we have
\begin{equation}
   8\pi\rho=\frac{b'(r_0)}{r_0^2}+\frac{1}{2}
   \frac{\partial\mu(r_0,l)}{\partial r}
   \frac{b'(r_0)-1}{r_0}>0
\end{equation}
since $\partial\mu(r_0,l)/\partial r <0$.  (Within
the four-dimensional framework, this reduces to the
usual $8\pi\rho =b'(r_0)/r_0^2$.)

To show that the NEC is met for the remaining null
vectors in the four-dimensional space, we first obtain
\begin{multline}
   \left.R_{00}+R_{22}\right|_{r=r_0}=
   \left.R_{00}+R_{33}\right|_{r=r_0}\\
   =-\frac{1}{2}\frac{d\Phi(r_0)}{dr}
   \frac{r_0b'(r_0)-b(r_0)}{r_0^2}+\frac{1}{2}
   \frac{r_0b'(r_0)-b(r_0)}{r_0^3}+\frac{1}{r_0^2}.
\end{multline}
Suppose we let $d\Phi(r_0)/dr=-2/r_0$ for now.  Then
$ \left.R_{00}+R_{22}\right|_{r=r_0}=0$ whenever
$b'(r_0)=1/3$.  So if $d\Phi(r_0)/dr<-2/r_0$, as
required, then we obtain our final condition
\begin{equation}
   b'(r_0)>\frac{1}{3}.
\end{equation}
To summarize, the NEC is met at and near the throat
for the four-dimensional null vector $(1,1,0,0)$;
the NEC is also met for the null vectors $(1,0,1,0)$
and $(1,0,0,1)$ whenever $b'(r_0)>1/3$.

This result can be easily generalized to the null
vector
\begin{equation*}
   (1,a,b,c),\quad 0\le a,b,c\le 1, \quad
   a^2+b^2+c^2=1
\end{equation*}
as follows:
\begin{multline*}
   G_{00}+a^2G_{11}+b^2G_{22}+c^2G_{33}\\
   =G_{00}+a^2G_{11}+b^2G_{22}+(1-a^2-b^2)G_{33}\\
   =R_{00}+\frac{1}{2}R+a^2(R_{11}-\frac{1}{2}R)
   +b^2(R_{22}-\frac{1}{2}R)\\+(1-a^2-b^2)(R_{33}
   -\frac{1}{2}R)\\
   =R_{00}+a^2R_{11}+b^2R_{22}+(1-a^2-b^2)R_{33}.
\end{multline*}
By writing
\begin{equation*}
   R_{00}=a^2R_{00}+b^2R_{00}+(1-a^2-b^2)R_{00},
\end{equation*}
we obtain by regrouping,
\begin{equation*}
   T_{\alpha\beta}k^{\alpha}k^{\beta}=
   a^2(R_{00}+R_{11})+b^2(R_{00}+R_{22})
     +(1-a^2-b^2)(R_{00}+R_{33})>0.
\end{equation*}
Such a rearrangement would work for any null vector
in the four-dimensional space.

Our final topic is the possibility that both $\Phi$
and $\lambda$ and hence $b$ are functions of $r$ and
$l$, already noted in the Introduction.  From the
standpoint of theoretical design, the forms $\Phi
=\Phi(r,l)$ and $b=b(r,l)$ may only be of marginal
interest.  However, since our goal is to determine
the effect of the fifth dimension, this case needs
to be considered, in spite of being a considerable
complication.

If $\Phi=\Phi(r)$ and $b=b(r)$ are replaced by $\Phi=
\Phi(r,l)$ and $b=b(r,l)$, respectively, then $d\theta^1$
is no longer equal to zero.  As a result, the 1-form
$\omega^4_{\phantom{0}1}$ in Eq. (\ref{E:omega}) becomes
\begin{equation*}
   \omega^4_{\phantom{0}1}=
   \frac{\partial\mu(r,l)}{\partial r}
      \left(1-\frac{b(r,l)}{r}\right)^{1/2}\theta^4
    -\frac{1}{2r}\left(1-\frac{b(r,l)}{r}\right)^{-1}
    \frac{\partial b(r,l)}{\partial l}e^{-\mu(r,l)}
    \,\theta^1.
\end{equation*}
Assuming that $b(r_0,l)=r_0$, $\omega^4_{\phantom{0}1}$
is undefined at the throat.  This carries over to the
component
\begin{multline*}
   R_{011}^{\phantom{000}0}=-\frac{1}{2}
   \frac{\partial\Phi(r,l)}{\partial r}\,
   \frac{r\partial b(r,l)/\partial r-b(r,l)}{r^2}\\
   +\frac{\partial^2\Phi(r,l)}{\partial r^2}
   \left(1-\frac{b(r,l)}{r}\right)
   +\left[\frac{\partial\Phi(r,l)}{\partial r}
   \right]^2\left(1-\frac{b(r,l)}{r}\right)\\
   +\frac{1}{2r}\frac{\partial\Phi(r,l)}{\partial l}
   \frac{\partial b(r,l)}{\partial l}
   \left(1-\frac{b(r,l)}{r}\right)^{-1}
   e^{-2\mu(r,l)}.
\end{multline*}
The component $ R_{144}^{\phantom{000}1}$ is also
undefined at the throat.  To determine the effect on
the Ricci scalar, we return to Eq. (\ref{E:Ricci})
and note that
\begin{multline*}
  \frac{1}{2}R=-R_{011}^{\phantom{000}0}
    -R_{022}^{\phantom{000}0}
    -R_{033}^{\phantom{000}0}
    -R_{044}^{\phantom{000}0}\\
  -R_{122}^{\phantom{000}1}
    -R_{133}^{\phantom{000}1}-R_{144}^{\phantom{000}1}
    -R_{233}^{\phantom{000}2}-R_{244}^{\phantom{000}2}
    -R_{344}^{\phantom{000}3}.
\end{multline*}
The undefined terms are sufficient to show that there
is a curvature singularity at the throat since $R$ is a 
scalar invariant.  So the redshift and shape functions 
must be functions of the radial coordinate only.


\section{Summary}
This paper explores the effect of an extra spatial
dimension on a Morris-Thorne wormhole.  It is proposed
that a natural choice for a model is the line element
\begin{equation*}\label{E:L4}
  ds^2=-e^{2\Phi(r)}dt^2+\frac{dr^2}{1-b(r)/r}+r^2
  (d\theta^2+\text{sin}^2\theta\,d\phi^2)+e^{2\mu(r,l)}dl^2,
\end{equation*}
where $l$ is the extra coordinate.  The goal is to show
that in the four-dimensional setting, the throat of the
wormhole can be threaded with ordinary matter, while the
unavoidable violation of the NEC can be attributed to the
existence of the extra dimension.

The precise conditions are: if
\begin{equation*}
   \frac{\partial\mu(r_0,l)}{\partial r}<-\frac{2}{r_0}
   \quad \text{and} \quad \frac{d\Phi(r_0)}{dr}=-A<
   \frac{\partial\mu(r_0,l)}{\partial r}
   <-\frac{2}{r_0},
\end{equation*}
then the NEC is met for the four-dimensional null vector
$(1,1,0,0)$ and violated for the five-dimensional null
vector $(1,0,0,0,1)$.  For the remaining four-dimensional
null vectors, we require the additional condition
$b'(r)>1/3$ at or near the throat in order to meet the
NEC.

For the fifth dimension, if the function $\mu(r,l)$ is
independent of $r$, then the extra dimension has no effect
on the conditions needed to sustain a regular
Morris-Thorne wormhole.  The functions $\Phi$ and $b$,
on the other hand, must be independent of $l$ to avoid
a curvature singularity at the throat.
\\
\\

\textbf{APPENDIX A\quad The curvature 2-forms}
\begin{equation*}
\Omega^0_{\phantom{0}1}=\left[\frac{1}{2}\frac{d\Phi(r)}{dr}
  \frac{rb'-b}{r^2}-\frac{d^2\Phi(r)}{dr^2}\left(1-\frac{b}{r}
     \right)\right.
     \left.-\left(\frac{d\Phi(r)}{dr}\right)^2\left(1
     -\frac{b}{r}\right)\right]\theta^0\wedge\theta^1
\end{equation*}
\begin{equation*}
   \Omega^0_{\phantom{0}2}=-\frac{1}{r}\frac{d\Phi(r)}{dr}
   \left(1-\frac{b}{r}\right)\theta^0\wedge\theta^2
\end{equation*}
\begin{equation*}
   \Omega^0_{\phantom{0}3}=-\frac{1}{r}\frac{d\Phi(r)}{dr}
   \left(1-\frac{b}{r}\right)\theta^0\wedge\theta^3
\end{equation*}
\begin{equation*}
   \Omega^0_{\phantom{0}4}=-\frac{d\Phi(r)}{dr}
   \frac{\partial\mu(r,l)}{\partial r}\left(1-\frac{b}{r}
   \right)\theta^0\wedge\theta^4
\end{equation*}
\begin{equation*}
  \Omega^1_{\phantom{0}2}=\frac{1}{2}\frac{rb'-b}{r^3}\,
     \theta^1\wedge\theta^2
\end{equation*}
\begin{equation*}
  \Omega^1_{\phantom{0}3}=\frac{1}{2}\frac{rb'-b}{r^3}\,
     \theta^1\wedge\theta^3
\end{equation*}
\begin{equation*}
   \Omega^1_{\phantom{0}4}=\left[-\frac{\partial^2\mu(r,l)}
   {\partial r^2}
   \left(1-\frac{b}{r}\right)+\frac{1}{2}
   \frac{\partial\mu(r,l)}{\partial r}\frac{rb'-b}{r^2}
   \right.
   \left.-\left(\frac{\partial\mu(r,l)}{\partial r}\right)^2
   \left(1-\frac{b}{r}\right)\right]\theta^1\wedge\theta^4
\end{equation*}
\begin{equation*}
   \Omega^2_{\phantom{0}3}=\frac{b}{r^3}\,\theta^2\wedge\theta^3
\end{equation*}
\begin{equation*}
   \Omega^2_{\phantom{0}4}=-\frac{1}{r}\frac{\partial\mu(r,l)}
   {\partial r}\left(1-\frac{b}{r}\right)\theta^2\wedge\theta^4
\end{equation*}
\begin{equation*}
   \Omega^3_{\phantom{0}4}=-\frac{1}{r}\frac{\partial\mu(r,l)}
   {\partial r}\left(1-\frac{b}{r}\right)\theta^3\wedge\theta^4
\end{equation*}
\\
\\
\textbf{APPENDIX B\quad The components of the Ricci tensor}
\begin{multline*}
  R_{00}=-\frac{1}{2}\frac{d\Phi(r)}{dr}\frac{rb'-b}{r^2}
  +\frac{d^2\Phi(r)}{dr^2}\left(1-\frac{b}{r}\right)\\+
  \left[\frac{d\Phi(r)}{dr}\right]^2\left(1-\frac{b}{r}\right)
  +\frac{2}{r}\frac{d\Phi(r)}{dr}\left(1-\frac{b}{r}\right)
    +\frac{d\Phi(r)}{dr}\frac{\partial\mu(r,l)}{\partial r}
  \left(1-\frac{b}{r}\right)
\end{multline*}
\begin{multline*}
   R_{11}=\frac{1}{2}\frac{d\Phi(r)}{dr}\frac{rb'-b}{r^2}
  -\frac{d^2\Phi(r)}{dr^2}\left(1-\frac{b}{r}\right)\\
  -\left[\frac{d\Phi(r)}{dr}\right]^2\left(1-\frac{b}{r}\right)
  +\frac{rb'-b}{r^3}-\frac{\partial^2\mu(r,l)}{\partial r^2}
  \left(1-\frac{b}{r}\right)\\
  +\frac{1}{2}\frac{\partial\mu(r,l)}{\partial r}
  \frac{rb'-b}{r^2}-\left[\frac{\partial\mu(r,l)}{\partial r}
  \right]^2\left(1-\frac{b}{r}\right)
\end{multline*}
\begin{equation*}
   R_{22}=R_{33}=-\frac{1}{r}\frac{d\Phi(r)}{dr}
   \left(1-\frac{b}{r}\right)+\frac{1}{2}\frac{rb'-b}{r^3}
   +\frac{b}{r^3}
   -\frac{1}{r}\frac{\partial\mu(r,l)}{\partial r}
   \left(1-\frac{b}{r}\right)
\end{equation*}
\begin{multline*}
  R_{44}=-\frac{d\Phi(r)}{dr}\frac{\partial\mu(r,l)}{\partial r}
  \left(1-\frac{b}{r}\right)-\frac{\partial^2\mu(r,l)}
  {\partial r^2}\left(1-\frac{b}{r}\right)\\
  +\frac{1}{2}\frac{\partial\mu(r,l)}{\partial r}\frac{rb'-b}{r^2}
  -\left[\frac{\partial\mu(r,l)}{\partial r}\right]^2
  \left(1-\frac{b}{r}\right)
  -\frac{2}{r}\frac{\partial\mu(r,l)}{\partial r}
  \left(1-\frac{b}{r}\right)
\end{multline*}

\end{document}